\documentstyle[preprint,oldlfont,eqsecnum,aps]{revtex}
\begin{document}
\draft
\preprint{U. C. Irvine Technical Report 95-8}
\title{
{\boldmath $B$} Factory Constraints on Isosinglet Down Quark Mixing,\\
and Predictions For Other {\boldmath $CP$} Violating Experiments}
\author{Dennis Silverman}
\address{Department of Physics and Astronomy, \\
University of California, Irvine \\
Irvine, CA 92717-4575 }
\maketitle

\begin{abstract}

In the main part of the paper we project forward to having $B$ factory
determinations of $\sin(2\beta)$ and $\sin(2\alpha)$, for which we
take several values.  First, we use a joint $\chi^2$ analysis of CKM
experiments to constrain CKM matrix elements in the standard model,
and experiments on the angles $\alpha$, $\beta$, and $\gamma$, and on
$x_s$ and null $CP$ asymmetries.  Then we invoke mixing to a new
isosinglet down quark (as in E$_6$) which induces FCNC that allow a
$Z^0$ mediated contribution to $B-\bar B$ mixing and which brings in
new phases.  We then repeat the $\chi^2$ analysis, now including
experimental constraints from FCNC as well, finding much larger ranges
of prediction for the $B$ factory.  We then add projected $B$ factory
results on $\sin(2\beta)$ and $\sin(2\alpha)$ and repeat both
analyses.  In $(\rho,\eta)$ and $(x_s,\sin{(\gamma)})$ plots for the
extra isosinglet down quark model, we find multiple regions that will
require experiments on $\sin{(\gamma)}$ and/or $x_s$ to decide between
and possibly to effectively bound out the extra down quark
contribution.
\end{abstract}

\section{Introduction}
In this paper we are interested in finding the ``reach'' of the $B$
factories in terms of determining the angles of the standard model
(SM) CKM matrix, and of limiting ``new physics'' contributions.
Within these future limits on both models, we then make predictions
for other experiments, including $\sin(\gamma)$, $x_s$, and the
asymmetry in $B_s-\bar B_s$ mixing, which is almost null in the SM.
In setting limits we use the method of a joint $\chi^2$ fit to all
constraining experiments.  The ``new physics'' class of models we use
are those with extra iso-singlet down quarks, where we take only one
new down quark as mixing significantly.  An example is E$_6$, where
there are two down quarks for each generation with only one up quark,
and of which we assume only one new iso-singlet down quark mixes
strongly.  This model has shown large possible effects in $B-\bar B$
mixing phases.  The ``reach'' of the $B$ factory in this model also
sets limits on the phases of the mixing angles to the new iso-singlet
down quark.  For different $\sin({2\alpha})$ we find multiple regions
that will require experiments on $\sin{(\gamma)}$ or $x_s$ to decide
between, and experiments on both could be required to effectively
bound out or to verify the model.  We also find a relatively small
$B_s - \bar{B}_s$ mixing asymmetry, even outside the standard model.

\section{Iso-singlet Down Quark Mixing Model}

Groups such as $E_6$ with extra SU$(2)_L$ singlet down quarks give
rise to flavor changing neutral currents (FCNC) through the mixing of
four or more down quarks \cite{shin,nirsilnp,nirsilpr,sil92}.  We use
the $4 \times 4$ down quark mixing matrix $V$ which diagonalizes the
initial down quarks ($d_{iL}^0$) to the mass eigenstates ($d_{jL}$) by
$d_{iL}^0 = V_{ij} d_{jL}$.  The flavor changing neutral currents we
have are \cite{nirsilpr,sil92} $-U_{ds} = V^*_{4d} V_{4s}$
, $-U_{sb} = V^*_{4s} V_{4b}$, and $-U_{bd} = V^*_{4b} V_{4d}$.  These
FCNC with tree level $Z^0$ mediated exchange may contribute part of
$B_d^0 - \bar{B_d^0}$ mixing and of $B_s^0 -
\bar{B_s^0}$ mixing, giving a range of non-zero values for the fourth
quark's mixing parameters.  $B_d^0 - \bar{B_d^0}$ mixing may occur by
the $b - \bar{d}$ quarks in a $\bar{B_d}$ annihilating to a virtual
$Z$ through a FCNC with amplitude $U_{db}\ $, and the virtual $Z$ then
creating $\bar{b} - d$ quarks through another FCNC, again with
amplitude $U_{db}$, which then becomes a $B_d$ meson.  If these are a
large contributor to the $B_d -\bar{B_d}$ mixing, they introduce three
new mixing angles and two new phases into the $CP$ violating $B$ decay
asymmetries.  The size of the contribution of the FCNC amplitude
$U_{db}$ as one side of the unitarity quadrangle is less than 0.1 of
the unit base $|V_{cd} V_{cb}|$ at the 1-$\sigma$ level, but we have found
\cite{shin,nirsilpr,sil92} that it can contribute, at present, as
large an amount to $B_d -\bar{B}_d$ mixing as does the standard model.
The new phases can appear in this mixing and give total
phases different from that of the standard model in $CP$
violating $B$ decay asymmetries\cite{nirsilpr,sil92,chosilfcnc,branco,lavoura}.

For $B_d - \bar{B}_d$ mixing with the four down quark induced $b-d$
coupling, $U_{db}$, we have \cite{chosilfcnc}
\begin{equation}
x_d = (2 G_F/3 \sqrt{2}) B_B f_B^2 m_B \eta_B \tau_B \left|
U_{std-db}^2 + U_{db}^2 \right|
\end{equation}
where with $y_t = m_t^2/m_W^2$
\begin{equation}
U^2_{std-db} \equiv  (\alpha/(4 \pi \sin^2{\theta_W})) y_t
f_2(y_t) (V_{td}^* V_{tb})^2,
\end{equation}
and $x_d = \Delta
m_{B_d}/\Gamma_{B_d} = \tau_{B_d} \Delta
m_{B_d}$.

The $CP$ violating decay asymmetries depend on the combined phases of
the $B^0_d-\bar{B}^0_d\;$ mixing and the $b$ quark decay amplitudes
into final states of definite $CP$.  Since we have found that $Z$
mediated FCNC processes may contribute significantly to
$B^0_d-\bar{B}^0_d\;$ mixing, the phases of $U_{db}\ $ would be
important.  Calling the singlet down quark $D$, to leading order the
mixing matrix elements to $D$ are $V_{tD} \approx s_{34}$, $V_{cD}
\approx s_{24} e^{-i\delta_{24} }$, and $V_{uD} \approx s_{14} e^{-i
\delta_{14} }$.  The FCNC amplitude $U_{db}$ to leading order in the new angles
is
\begin{equation}
U_{db} =
-s_{34} (s_{34} V^*_{td} + s_{14} e^{-i \delta_{14} } -s_{24} e^{-i
\delta_{24} } s_{12}).
\end{equation}
where $V_{td} \approx (s_{12} s_{23} - s_{13} e^{i\delta_{13}})$, and
$V_{ub} = s_{13} e^{-i\delta_{13}}$.

\section{Joint Chi-squared Analysis for CKM and FCNC Experiments}

FCNC experiments put limits on the new mixing angles and
constrain the possibility of new physics contributing to the $B_d^0 -
\bar{B_d^0}$ and $B_s^0 - \bar{B_s^0}$ mixing.  Here we analyze
jointly all constraints on the $4 \times 4$ mixing matrix obtained by
assuming only one of the SU$(2)_L$ singlet down quarks mixes
appreciably\cite{nirsilpr}.  We use the eight experiments for the $3
\times 3$ CKM sub-matrix elements \cite{chosilckm}, which include
those on the five matrix elements $V_{ud}, V_{cd}, V_{us}, V_{ub},
V_{cb}$ of the $u$ and $c$ quark rows, and, in the neutral $K$
system\cite{detail}, include $|\epsilon|$ and $K_L \to \mu \mu$, and also
$B_d -
\bar{B}_d$ mixing.  For studying FCNC, we add \cite{chosilfcnc}
the $B \to \mu \mu X$ bound (which constrains $b \to d$ and $b \to
s$), $K^+ \to \pi^+ \nu \bar{\nu}$ \cite{lavoura,kpiexpt,chiskpi}
and $Z^0 \to b \bar{b}$ \cite{lavoura}
(which directly constrains the $V_{4b}$
mixing element).  FCNC experiments will bound the three
amplitudes $U_{ds}$, $U_{sb}$, and $U_{bd}$ which contain three new
mixing angles and three phases.  We use the newly indicated mass of
the top quark as $m_t = 174$ GeV.

In maximum likelihood correlation plots, we use for axes two output
quantities which are dependent on the angles, such as $\rho$ and
$\eta$, and for each possible bin with given values for these,
we search through the nine dimensional angular data set of the $4
\times 4$ down quark mixing angles, finding all sets which give results in
the bin, and then putting into that bin the minimum $\chi^2$ among
them.  To present the results, we then draw contours at several
$\chi^2$ in this plane corresponding to given confidence levels.

\section{Constraints on the Standard Model CKM Matrix at Present, and
After the B Factory}

We first analyze the standard model using the present constraints on
the eight CKM related experiments, and then repeat the analysis using
the projected constraints from the $B$ factory\cite{Buras} which will
give values for $\sin(2\beta)$ and $\sin(2\alpha)$.  In the following,
we will find and take $\sin(2\beta) = 0.62$ as the center of the
current range with its projected $B$ factory errors of $\pm 0.06$
\cite{porter}, and vary $\sin(2\alpha)$ from $-1.0$ to 1.0, using the
projected $B$ factory errors of $\pm 0.08$.

In Fig.~\ref{rho-etaSM} is shown the $(\rho,\eta)$ plot for the
standard model with contours at $\chi^2$ which correspond to confidence
levels (CL) that are the same as the CL for 1, 2, and 3-$\sigma$
limits.  Fig.~\ref{rho-etaSM} shows large regions for the present CKM
constraints, and small regions for the projected $B$ factory
results, where we have taken the cases $\sin(2\alpha)= 1, 0,$ and
$-1$, which appear from left to right, respectively.

In Fig.~\ref{sa-sbSM} is shown the $(\sin(2\alpha),\sin(2\beta))$ plot
for the standard model, for the same cases as in Fig. 1.  The
nearly horizontal contours are the present constraints, and the small
circular contours are those for the $B$ factory cases
$\sin{(2\alpha)} = -1, 0$, and 1, centered about their appropriate
$\sin{(2\alpha)}$ values.

In Fig.~\ref{xs-sgSM} is shown the $(x_s,\sin{(\gamma)})$ plot for the
standard model with (a) present data, and (b) for the $B$
factory cases $\sin{(2\alpha)} =
1, 0, -1$ from left to right.  $x_s$ is determined here from $x_s =
1.2 x_d (|V_{ts}|/|V_{td}|)^2$.  The largest errors arise from the
uncertainty in $|V_{td}|$, since we have not assumed any improvement
in the present $20\%$ uncertainty in $\sqrt{B_B} f_B$ (which relates
$V_{td}$ to $x_d$) from lattice calculations\cite{latticefB}.  The $B$
factory in the SM constructs a rigid triangle from the knowledge of
$\alpha$ and $\beta$, and removes this uncertainty in $\gamma$ and
$x_s$ in the future.  A cautionary note for experiments emerges from
this plot, namely that $\sin(\gamma)$ is close to one (0.8 to 1.0) for
the 1-$\sigma$ contour, and high accuracy on $\sin(\gamma)$ will be
needed to add new information to the standard model.
At 1-$\sigma$ the range of $x_s$ in the standard model is
from 11 to 24.  It is clear that the choices of $\sin(2\alpha)$ cases
gives distinct ranges for $x_s$.  Using $x_s$ to agree with the range
given by a $\sin{(2\alpha)}$ measurement will be a good test of the
standard model.

\section{Constraints on the Four Down Quark Model at Present, and After the
{\boldmath $B$} Factory Results}

Here we also project forward to having results on $\sin{(2\alpha)}$
and $\sin{(2\beta)}$ from the $B$ factories, and show how there will
be stronger limits on the new phases of FCNC couplings than from
present data.  In the four down quark model we use
``$\sin{(2\alpha)}$'' and ``$\sin{(2\beta)}$'' to denote results of
the appropriate $B_d$ decay $CP$ violating asymmetries, but since the
mixing amplitude is a superposition, the experimental results are not
directly related to angles in a triangle in this model.  The
asymmetries with FCNC contributions included are
\begin{equation}
\sin{(2\beta)} \equiv A_{B^0_d \to \Psi K^0_s} =
{\rm Im} \left[ \frac{(U^2_{std-db} + U^2_{db})}{|U^2_{std-db} + U^2_{db}|}
\frac{(V^*_{cb} V_{cs})}{(V^*_{cb} V_{cs})^*} \right]
\end{equation}
\begin{equation}
\sin{(2\alpha)} \equiv -A_{B^0_d \to \pi^+ \pi^-} =
-{\rm Im} \left[ \frac{(U^2_{std-db} + U^2_{db})}{|U^2_{std-db} + U^2_{db}|}
\frac{(V^*_{ub} V_{ud})}{(V^*_{ub} V_{ud})^*} \right]
\end{equation}
with $U_{std-db}$ defined in Eqn. (2.2).

We analyze all of these constraints together using a joint $\chi^2$
for fitting all of the thirteen experiments in the nine parameter angle
space of the $4 \times 4$ mixing matrix.  We include both the standard
model and FCNC contributions through effective
Hamiltonians\cite{chosilfcnc}. We then make maximum likelihood plots which
include ($\sin{(2\alpha)}$, $\sin{(2\beta)}$), ($\rho$, $\eta$),
($x_s$, $\sin{\gamma}$), and those involving the FCNC amplitudes
$U_{db}$ and $U_{sb}$ (not shown).

The corresponding plots for the four down quark model are shown for
present data and for projected $B$ factory data
in the following figures.  In the figures, we show
$\chi^2$ contour plots with confidence levels (CL) at values
equivalent to 1-$\sigma$ and at 90\% CL (1.64$\sigma$) for present
data, and for projected $B$ factory results.
Again, for results with the $B$ factories, we use the example of the
most likely $\sin{(2\beta)} = 0.62$ with $B$ factory errors of $\pm
0.06$, and errors of $\pm 0.08$ on $\sin{(2\alpha)}$.

In Fig.~\ref{rho-eta4q} we have plotted the $\chi^2$ contours for the
location of the vertex of $V_{ub}^*V_{ud}/|V_{cb}V_{cs}| \equiv \rho +
i\eta$ (even for the four down quark quadrangle case).  We note that
in contrast to the standard model, in Fig.~\ref{rho-eta4q}a the
presently allowed contours in the four down quark model go down to
$\eta = 0$ at the 90\% CL, which can result from the FCNC with its
phases in $U_{db}$ causing the known $CP$ violation.  In
Fig.~\ref{rho-eta4q}b,c and d we show the $B$ factory cases of
$\sin{(2\alpha)} = -1, 0$ and 1, respectively, with contours at
1-$\sigma$ and at 90\% CL.  The existence of several regions requires
that extra experiments in $\sin{(\gamma)}$ or $x_s$ will also be
needed to verify or to bound out the extra down quark mixing model.
The larger contours at 1-$\sigma$ roughly agree with those of the
standard model in Fig. 1.  In our $\chi^2$ we have
used\cite{chosilckm,chosilfcnc} $|V_{ub}| = 0.071 \pm 0.013$
consistent with adjusting $\kappa$ in the Isgur-Wise model to fit the
spectra, and using the spread of model results to determine a
$\sigma$.  In any case, we can consider this accuracy as obtainable
in the future, following ref.~\cite{Buras}.
Use of the conservative bound of $0.08 \pm 0.03$ used by
others still results in multiple regions.

The $(\sin{(2\alpha)},\sin{(2\beta)})$ $\chi^2$ contour plot for the
four down quark model (not shown) shows that all values of
$\sin{(2\beta)}$ and $\sin{(2\alpha)}$ are individually allowed at
1-$\sigma$, and most pairs of values are allowed at 1-$\sigma$.  This
is a much broader allowed region in $\sin{(2\beta)}$ than the standard
model result from present data in Fig. 2.  The allowed 1,2 and
3-$\sigma$ contours in the ($\sin{(2\alpha)}$, $\sin{(2\beta)}$) plot
for the cases of the $B$ factory results with the four down quark
model are very similar to the SM results shown in Fig. 2.

In terms of other experiments, the $(x_s,\sin{(\gamma)}$) plot
for the four down quark model is shown in Fig.~\ref{xs-sg4q}a with the
allowed region from present data, with 1-$\sigma$ and 90\% CL contours.
This allows all values of $\sin{(\gamma)}$ at the 1-$\sigma$ CL at present,
and at 1-$\sigma$ constrains $x_s$ to lie between 8 and 25.  In the
four down quark model, what we mean by ``$\sin{(\gamma)}$'' is the
result of the experiments which would give this variable in the SM
\cite{Kayser}.  Here, the four down quark model involves more
complicated amplitudes, and is not simply $\sin{(\delta_{13})}$
\begin{equation}
\sin(\gamma) = {\rm Im}   \left[ \frac{(U^2_{std-bs} + U^2_{bs})}
{|U^2_{std-bs} + U^2_{bs}|}
\frac{(V^*_{ub} V_{cs})}{|V_{ub} V_{cs})|} \right],
\end{equation}
\begin{equation}
x_s = 1.2 x_d \frac{|U^2_{std-bs}+U^2_{bs}|}{|U^2_{std-db}+U^2_{db}|},
\end{equation}
where
\begin{equation}
U^2_{std-bs} = (\alpha/(4 \pi \sin{\theta_W}^2))y_t f_2(y_t)
(V^*_{tb}V_{ts})^2.
\end{equation}

In Figs.~\ref{xs-sg4q}b, c and d are shown the cases $\sin{(2\alpha)}
= -1, 0$, and 1, respectively, at 1-$\sigma$ and at 90\% CL.  They
reflect the same regions that appeared in the $(\rho,\eta)$ plots,
Figs.~\ref{rho-eta4q}b, c, and d.  The resemblance is increased if we
recall that roughly $\sin{(\gamma)}
\approx \eta$, and also that $x_s \propto 1/|V_{td}|^2$ where $|V_{td}|$
is the distance from the $\rho=1, \eta=0$ point.  We see that
experiments on $\sin{(\gamma)}$ and $x_s$ are necessary to resolve the
possible regions allowed by the four down quark model.  For the case
of $\sin{(2\alpha)}=-1$, the allowed values of $\sin{(\gamma)}$
in Fig.~\ref{xs-sg4q}b
are smaller than those for the standard model in Fig.~3a.

The asymmetry $A_{B_s}$ in $B_s$ mixing in the standard model with the
leading decay process of $b \to c \bar c s$ has no significant phase
from the decay or from the mixing which is proportional to
$V_{ts}^2$.  The vanishing of this asymmetry is a test of the standard
model\cite{nirsilnp}, and a non-zero value can result from a ``new
physics'' model.  With the FCNC, the new result is
\begin{equation}
A_{B_s}  =
{\rm Im} \left[ \frac{(U^2_{std-bs} + U^2_{bs})}{|U^2_{std-bs} + U^2_{bs}|}
\frac{(V^*_{cb} V_{cs})}{(V^*_{cb} V_{cs})^*} \right]
\end{equation}
The extent of the non-zero value of $A_{B_s}$ in the four down quark
model is shown in Fig.~\ref{xs-Bs4q} from present data with contours
at 1, 2 and 3-$\sigma$.  Plots for the $B$ factory cases
(not shown) are similar.  We note that it is bounded to be rather
small from present data at 1-$\sigma$, i.e. less than 0.06, and less than
0.32 at 2-$\sigma$.

We compared the limits on the four down quark FCNC amplitude
$|U_{db}|$ versus the standard model amplitude $|U_{std-db}|$ for
$B_d^0 - \bar{B}_d^0$ mixing , at present and after the $B$ factory
results.  At present the constraints are such that $|U_{db}|$ can go
from zero up to as large as the magnitude of $|U_{std-db}|$ at 2-$\sigma$
\cite{chosilfcnc}.  For the sample $B$ factory
results, the $|U_{db}|$ range is only somewhat more restricted.  The total
phase of $B_d^0 - \bar{B}_d^0$ mixing is closely restricted, however,
to the same range as the standard model amplitude.

The 90\% CL limits on the three new quark mixing elements
$|V_{4d}|$, $|V_{4s}|$, and $|V_{4b}|$ are roughly equal to the mixing
angles to the fourth down quark $\theta_{14}$, $\theta_{24}$ and
$\theta_{34}$, respectively.  They are bounded by 0.06, 0.05, and
0.14, respectively.

\section{Conclusions}

The main conclusion with the four down quark model for the $B$ factory
cases is that there are multifold allowed regions as shown in the
$(\rho,\eta)$ plot and the $(x_s,\sin{(\gamma)})$ plot.  This will
require additional experiments on $x_s$ and $\sin{(\gamma)}$ to well
define the four down quark model results, and eventually to verify or
bound out the relevance of the model here.

We note that in the four down quark model the $\eta \propto $
Im$(V_{ub}^*)$ range can reach zero, which is quite different than in
the standard model.  This is because the other phases can account for
$CP$ violation.

We have also found that the present range of $x_s$ at 1-$\sigma$ is
from 11 to 24 in the standard model, and from 8 to 25 in the four down
quark model at 1-$\sigma$.  The $\sin{(\gamma)}$ range is from 0.8 to 1.0
in the SM at 1-$\sigma$, and completely undetermined in the four down quark
model at 1-$\sigma$.

Finally, the range of the $B_s$ asymmetry which almost vanishes in the
standard model, is found to range from zero up to 0.06 at 1-$\sigma$ in
the four down quark model.  Although its presence is a signal against
the standard model, it may be small in new physics models, as it is in
this one, and thus hard to detect.

\acknowledgements

This research was supported in part by the U.S. Department of Energy
under Contract No. DE-FG0391ER40679.  We acknowledge the hospitality of the
Aspen Center for Physics.

\begin{figure}
\caption{The $(\rho,\eta)$ plot for the standard model, showing the 1, 2,
and 3-$\sigma$ contours, for the present data (large contours) and for
projected $B$ factory results (smaller circular contours) at
$\sin{(2\alpha)}= 1,0$, and $-1$ from left to right.}
\label{rho-etaSM}
\end{figure}

\begin{figure}
\caption{
The ($\sin{(2\alpha)}$, $\sin{(2\beta)}$) plot for the standard model
at 1, 2, and 3-$\sigma$ with present data (nearly horizontal contours),
and with the sample results of the $B$ factories (almost circular
contours), for $\sin{(2\alpha)} = 1, 0$, and $-1$ from left to right.
\label{sa-sbSM} }
\end{figure}

\begin{figure}
\caption{
The ($x_s$, $\sin{\gamma}$) plots are shown for the standard model
with:  (a) present limits; and (b) sample results for the $B$
factories for $\sin{(2\alpha)} = 1, 0$, and $-1$ from left to right.
\label{xs-sgSM} }
\end{figure}

\begin{figure}
\caption{The $(\rho,\eta)$ plots for the four down quark model from:
(a) present data, and for $B$ factory cases for values of
$\sin{(2\alpha)}$ as labeled.  Contours are at
1-$\sigma$ and at 90\% CL.
\label{rho-eta4q} }
\end{figure}

\begin{figure}
\caption{The $(x_s,\sin{(\gamma)})$ plots for the four down quark model from
(a) present data, and (b, c, and d) for $B$ factory cases for values of
$\sin{(2\alpha)}$ as labeled.  Contours are the same as in Fig.~4.
\label{xs-sg4q} }
\end{figure}

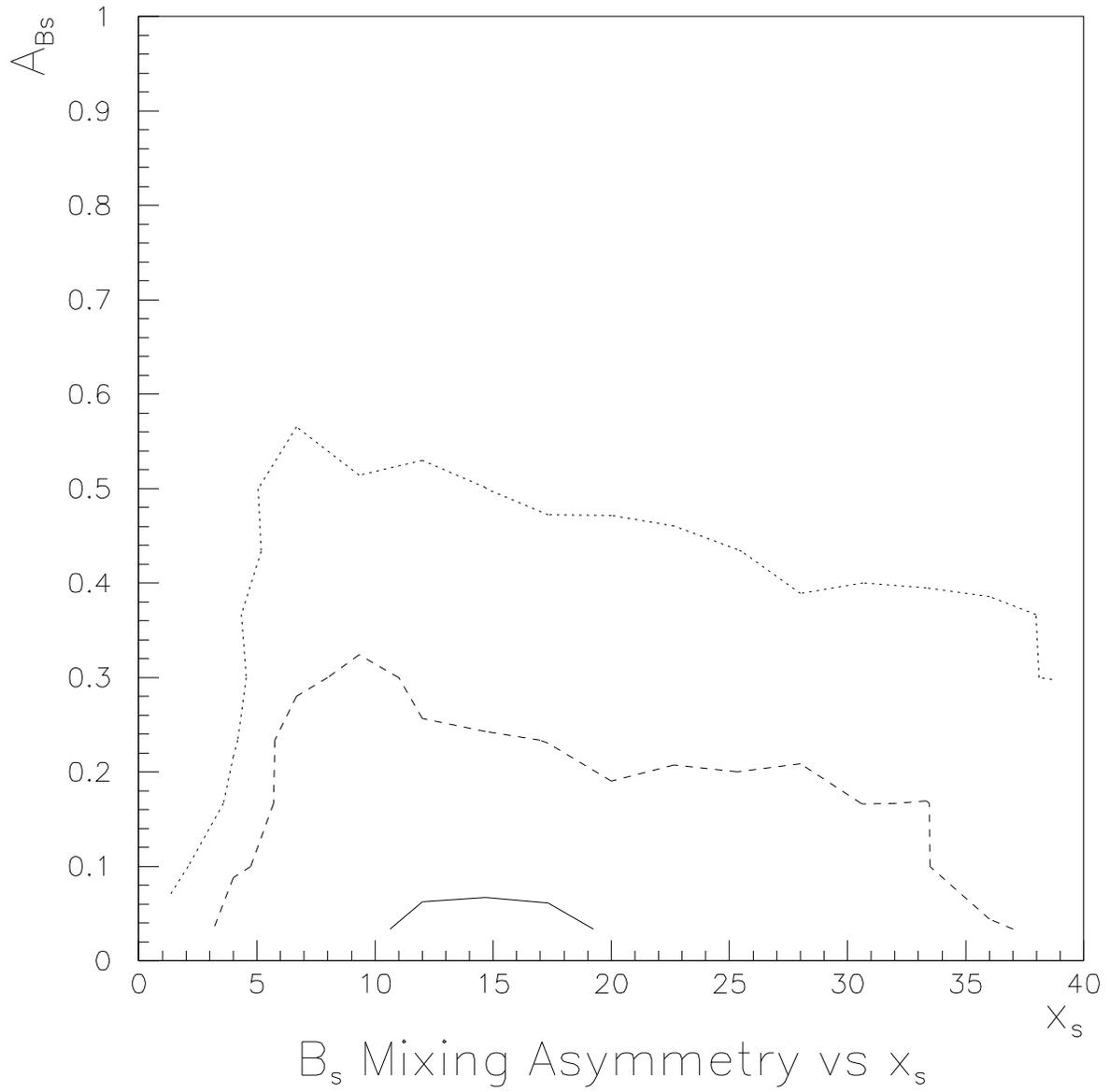
\begin{figure}
\caption{The $(x_s,A_{B_s})$ plot for the $B_s$ asymmetry $A_{B_s}$ in
the four down quark model for present data, with contours at
1, 2 and 3-$\sigma$.
\label{xs-Bs4q} }
\end{figure}


\begin{references}
\bibitem{chiskpi} With ${\rm BR}(K^+ \to \pi^+ \nu {\bar \nu}) < 5.2
\times 10^{-9}$
at 90\% CL, and ignoring the much smaller standard model contribution,
we have the $Z^0$ exchange FCNC bound in $\Delta \chi^2 = (|U_{ds}|^2/
6.5 \times 10^{-9})^2$ for this experiment.
\end{references}
\end{document}